\documentstyle[aps,prl,multicol,bm]{revtex}

\begin{document}
\newcommand{\tr}{\mathop{\mathrm{tr}}\nolimits}
\newcommand{\wideeqtop}{%
\vskip-3.5mm
\noindent
  \vrule width3.4in height.2pt depth.2pt %
  \vrule depth0em height2mm \hfill
\vspace*{1.5mm}
}
\newcommand{\wideeqend}{%
\vskip1mm
\indent
  \hfill\vrule depth2mm height0pt %
  \vrule width3.4in height.2pt depth.2pt
\vskip0mm
}
\newcommand{\referencetop}{%
\begin{center}
\vrule width83mm height.2pt depth.2pt\\
\medskip
\end{center}
}

\preprint{WU-HEP-00-8}
\title{%
Stochastic limit approximation for rapidly decaying systems}
\author{%
Gen Kimura,%
\thanks{Email: gen@hep.phys.waseda.ac.jp}
Kazuya Yuasa,%
\thanks{JSPS Research Fellow. Email: yuasa@hep.phys.waseda.ac.jp}
and
Kentaro Imafuku%
\thanks{Email: imafuku@mn.waseda.ac.jp}}
\address{%
Department of Physics, Waseda University, Tokyo 169-8555, Japan}
\date{December 7, 2000}
\maketitle
\begin{abstract}
The stochastic limit approximation method for ``rapid'' decay is
presented, where the damping rate $\gamma$ is comparable to the
system frequency ${\mit\Omega}$, i.e., $\gamma\sim{\mit\Omega}$,
whereas the usual stochastic limit approximation is applied only
to the weak damping situation $\gamma\ll{\mit\Omega}$.
The key formulas for rapid decay are very similar to those for
weak damping, but the dynamics are quite different.
From a microscopic Hamiltonian, the spin-boson model, a Bloch
equation containing two independent time scales is derived.
This is a useful method to extract the minimal dissipative
dynamics at high temperature $k_BT\gg\hbar{\mit\Omega}$ and the
master equations obtained are of the Lindblad form unlike that of
Caldeira and Leggett.
The validity of the method is confirmed by comparing the master
equation derived through this method with the exact one.
\end{abstract}

\begin{multicols}{2}\narrowtext
One of the most attractive issues in fundamental physics is to
understand dissipative dynamics from a microscopic point of view.
This is not a trivial question since quantum theory is designed
for closed systems and has time-reversal symmetry.
A standard approach to this problem is to deal with the
(dissipative) system of interest together with its surroundings,
the ``environment,'' which has infinite degrees of
freedom~\cite{ref:QuantumDissipativeSystems,ref:QuantumNoise,%
ref:FeynmanVernon,ref:Ullersma}.
The total system---``system'' + ``environment''---is then a
closed system and is treated in a quantum-mechanical way.
Caldeira and Leggett succeeded in formulating the ``quantum
Brownian motion'' by making use of the path-integral
method~\cite{ref:CaldeiraLeggettPhysica,%
ref:GrabertSchrammIngold,ref:HuPazQBM}, and recent developments
in technology have revealed the validity of this approach in the
field of quantum optics~\cite{ref:QuantumOpticsMilburn}.

One of the powerful methods of deriving dissipative dynamics from
the total system approach is the stochastic limit approximation
method established by Accardi
\textit{et~al.\/}~\cite{ref:SLA,ref:SLAinSpinBoson}.
In quantum optics, for example, one is often interested in the
decay of an excited atom in the radiation field.
The total Hamiltonian reads
\begin{equation}
\label{eqn:WeakDampingHamiltonian}
H_{\mathrm tot}=H_S+\lambda V+H_B,
\end{equation}
$H_S$ and $H_B$ being the Hamiltonians of the atom and the
radiation field, respectively.
Since the electromagnetic interaction $\lambda V$ is weak
($\lambda\ll1$), the weak-coupling limit $\lambda\to0$ gives a
good approximation.
Simultaneously making the coarse graining in time
$t=\tau/\lambda^2$ ($\lambda\to0$ with $\tau$ fixed), the
stochastic limit approximation extracts the minimal (but notable)
dissipative dynamics~\cite{ref:VanHove,ref:Spohn,%
ref:PascazioFacchiNonexponential-VanHoveLimit}.

However, the usual stochastic limit approximation applied to the
Hamiltonian (\ref{eqn:WeakDampingHamiltonian}) gives a rather
restricted dynamics, i.e., \textit{weak damping}.
The lowest-order contribution of the interaction to the damping
rate $\gamma$ is proportional to $\lambda^2$, and it is much
smaller than the characteristic frequency of the system
${\mit\Omega}$, i.e., $\gamma\ll{\mit\Omega}$, which is the case
of quantum optics.
For faster decay $\gamma\sim{\mit\Omega}$, which we call
\textit{rapid decay}, a separate treatment is needed.

The rapidly decaying dynamics is also of physical interest,
because it includes the case of quantum Brownian
motion~\cite{ref:QuantumNoise}.
It contains a richer variety of transient dynamics in contrast to
the situation of weak damping, where frequent oscillations of the
system allow the ``rotating-wave approximation,'' which
simplifies the dynamics
considerably~\cite{ref:ExactDiagonalization}.

In this paper, we explore the stochastic limit approximation for
rapid decay.
The total Hamiltonian is~\cite{ref:QuantumNoise,ref:Spohn}
\begin{equation}
\label{eqn:RapidDecayHamiltonian}
H_{\mathrm tot}=\lambda^2H_S+\lambda V+H_B.
\end{equation}
The system frequency ${\mit\Omega}$ is, in turn, of order
$O(\lambda^2)$, and the resultant dissipative dynamics exhibits
rapid decay $\gamma\sim{\mit\Omega}$.
It is shown that the key formulas are different from but similar
to those obtained in the usual stochastic limit approximation for
weak damping.
This brings us with a Bloch equation with two independent time
scales from a microscopic Hamiltonian, i.e., the spin-boson
model.
Furthermore, the master equations derived through this method are
shown to be of the Lindblad
form~\cite{ref:LindbladCommunMathPhys1976}, which ensures the
conservation and positivity of probability,
unlike~\cite{ref:QuantumNoise,ref:NonLindblad,%
ref:NonRotatingWaveMasterEq} that of Caldeira and
Leggett~\cite{ref:CaldeiraLeggettPhysica}.

For the sake of simplicity, we shall focus on the following
spin-boson model~\cite{ref:SpinBoson}.
The extension of the spin system to a general $N$-level system is
straightforward.
The total Hamiltonian is given by
Eq.~(\ref{eqn:RapidDecayHamiltonian}) with
\begin{mathletters}
\label{eqn:SpinBosonModel}
\begin{equation}
\label{eqn:BathHamiltonian}
H_S
=\frac{\varepsilon}{2}\sigma_z+\frac{\mit\Delta}{2}\sigma_x,\quad
H_B=\int dk\,\hbar\omega_ka_k^\dagger a_k,
\end{equation}
\begin{equation}
V=i\hbar\sigma_z\int dk\left(g_ka_k-g_k^*a_k^\dagger\right).
\end{equation}
\end{mathletters}
The spin system $H_S$, which has two energy eigenstates
\begin{equation}
H_S|\pm\rangle=\pm\frac{1}{2}\hbar{\mit\Omega}_0|\pm\rangle,\quad
\hbar{\mit\Omega}_0=\sqrt{\varepsilon^2+{\mit\Delta}^2},
\end{equation}
interacts with the boson system $H_B$ through the linear coupling
interaction $V$.
The particular choice of the coupling here is not essential
except the linearity;
$g_k$ is the coupling constant for the boson of mode $k$, and
throughout this paper it is assumed, as usual, that, before time
$t=0$, the two systems are uncorrelated and the boson system is
in the thermal equilibrium state at temperature $T$.
The initial state of the total system $\rho_{\mathrm tot}(0)$ is
thus given by
\begin{mathletters}
\begin{equation}
\rho_{\mathrm tot}(0)=\rho_S\otimes\rho_B,
\end{equation}
\begin{equation}
\rho_B=\frac{1}{Z}e^{-H_B/k_BT},\quad
Z=\tr_Be^{-H_B/k_BT},
\end{equation}
\end{mathletters}
where $k_B$ is the Boltzmann constant.
This model was discussed by the usual stochastic limit
approximation in Ref.~\cite{ref:SLAinSpinBoson} but with the
total Hamiltonian~(\ref{eqn:WeakDampingHamiltonian}).

With $H_0=\lambda^2H_S+H_B$ being the free part of the
Hamiltonian, the time-evolution operator in the interaction
picture, $U_I(t)$, satisfies the Schr\"odinger equation
\begin{mathletters}
\label{eqn:SchroedingerEq}
\begin{equation}
\frac{d}{dt}U_I(t)
=-\frac{i}{\hbar}\lambda V_I(t)U_I(t),\quad
U_I(0)=1,
\end{equation}
\begin{eqnarray}
V_I(t)&=&e^{iH_0t/\hbar}Ve^{-iH_0t/\hbar}\nonumber\\
&=&i\hbar\sum_\alpha\,\Bigl(
D_\alpha A_\alpha(t)-D_\alpha^\dagger A_\alpha^\dagger(t)
\Bigr).
\end{eqnarray}
\end{mathletters}
Here $D_\alpha\,(\alpha=\{+,-,0\})$ are the basic operators of
the spin system,
\begin{equation}
D_\pm=|\pm\rangle\langle\mp|,\quad
D_0=|+\rangle\langle+|-|-\rangle\langle-|,
\end{equation}
and $A_\alpha(t)$ the operators of the Bose field,
\begin{equation}
A_\alpha(t)=C_\alpha\int dk\,g_ka_k
e^{-i(\omega_k-\lambda^2\omega_\alpha)t},
\end{equation}
with $C_+=\langle+|\sigma_z|-\rangle%
={\mit\Delta}/\hbar{\mit\Omega}_0=C_-^*$,
$C_0=\langle+|\sigma_z|+\rangle=-\langle-|\sigma_z|-\rangle%
=\varepsilon/\hbar{\mit\Omega}_0$,
$\omega_\pm=\pm{\mit\Omega}_0$, and $\omega_0=0$.
Coarse graining in time is made in the Schr\"odinger
equation~(\ref{eqn:SchroedingerEq}):
Changing the time scale from the microscopic time $t$ to the
macroscopic one $\tau=\lambda^2t$,
\begin{eqnarray}
&&\frac{d}{d\tau}U_I(\tau/\lambda^2)\nonumber\\
&&\quad=\sum_\alpha\left(
D_\alpha\frac{1}{\lambda}A_\alpha(\tau/\lambda^2)
-D_\alpha^\dagger
\frac{1}{\lambda}A_\alpha^\dagger(\tau/\lambda^2)
\right)U_I(\tau/\lambda^2),\nonumber\\
\end{eqnarray}
we take the stochastic limit $\lambda\to0$ and obtain
\begin{equation}
\label{eqn:SchroedingerEqInStochasticLimit}
\frac{d}{d\tau}{\mathcal U}_I(\tau)
=\sum_\alpha\,\Bigl(
D_\alpha b_\alpha(\tau)-D_\alpha^\dagger b_\alpha^\dagger(\tau)
\Bigr)\,{\mathcal U}_I(\tau).
\end{equation}
Note that at the present stage,
Eq.~(\ref{eqn:SchroedingerEqInStochasticLimit}) should not be
regarded as mathematically fully justified.
A more rigorous analysis requires a prescription to handle normal
ordering~\cite{ref:SLA,ref:SLAinSpinBoson}.
We shall see later that a consistent procedure of normal ordering
can be obtained for the Heisenberg equations of the dressed spin
system operators.

Let us compute the correlation functions of the operators of the
Bose field $b_\alpha(\tau)$ in the thermal state $\rho_B$.
These operators play the role of the ``quantum noises'' in the
macroscopic time scale.
First notice the basic formula
\begin{eqnarray}
\label{eqn:Formula}
&&\lim_{\lambda\to0}\int_{-\infty}^\infty d\tau\,F(\tau)
\int_{-\infty}^\infty\frac{d\omega}{2\pi}G(\omega)
\frac{1}{\lambda^2}
e^{-i(\omega-\lambda^2\tilde{\omega})\tau/\lambda^2}\nonumber\\
&&\quad
=\lim_{\lambda\to0}\int_{-\infty}^\infty dt\,F(\lambda^2t)
\hat{G}(t)e^{i\lambda^2\tilde{\omega}t}\nonumber\\
&&\quad
=F(0)\lim_{\lambda\to0}G(\lambda^2\tilde{\omega}).
\end{eqnarray}
$\hat{G}(t)$ is the Fourier transform of $G(\omega)$.
The correlation functions in the stochastic limit read
\begin{mathletters}
\label{eqn:Correlation}
\begin{eqnarray}
&&\langle
b_\alpha(\tau)b_{\alpha'}^\dagger(\tau')
\rangle_B\nonumber\\
&&\qquad=\lim_{\lambda\to0}
{\mit\Gamma}_{\alpha\alpha'}^+(\lambda^2\omega_{\alpha\alpha'})
e^{i(\omega_\alpha-\omega_{\alpha'})\tau}\delta(\tau-\tau'),
\end{eqnarray}
\begin{eqnarray}
&&\langle
b_\alpha^\dagger(\tau)b_{\alpha'}(\tau')
\rangle_B\nonumber\\
&&\qquad
=\lim_{\lambda\to0}\Bigl(
{\mit\Gamma}_{\alpha\alpha'}^-(\lambda^2\omega_{\alpha\alpha'})
\Bigr)^*
e^{-i(\omega_\alpha-\omega_{\alpha'})\tau}\delta(\tau-\tau')
\end{eqnarray}
\end{mathletters}
with the spectral functions
\begin{mathletters}
\begin{equation}
{\mit\Gamma}_{\alpha\alpha'}^+(\omega)
=\Bigl(1+N(\omega)\Bigr)\,{\mit\Gamma}_{\alpha\alpha'}(\omega),
\end{equation}
\begin{equation}
{\mit\Gamma}_{\alpha\alpha'}^-(\omega)
=N(\omega){\mit\Gamma}_{\alpha\alpha'}(\omega),
\end{equation}
\begin{equation}
\label{eqn:SpectralFunction}
{\mit\Gamma}_{\alpha\alpha'}(\omega)
=C_\alpha C_{\alpha'}^*{\mit\Gamma}(\omega)
=2\pi C_\alpha C_{\alpha'}^*
\int dk\,|g_k|^2\delta(\omega_k-\omega),
\end{equation}
\end{mathletters}
the Bose-Einstein distribution function
\begin{equation}
N(\omega)=\frac{1}{e^{\hbar\omega/k_BT}-1},
\end{equation}
and the shorthand notation
$\omega_{\alpha\alpha'}=(\omega_\alpha+\omega_{\alpha'})/2$.
The correlation time of the operators of the Bose field
$b_\alpha(\tau)$ is negligibly small in the macroscopic time,
which makes it reasonable to call them quantum noises.

The factor $\lim_{\lambda\to0}{\mit\Gamma}_{\alpha\alpha'}^\pm%
(\lambda^2\omega_{\alpha\alpha'})$ is sensitive to the form of
the spectral function ${\mit\Gamma}_{\alpha\alpha'}(\omega)$ for
small positive $\omega$.
[For $\omega<0$, ${\mit\Gamma}_{\alpha\alpha'}(\omega)=0$ by the
definition in Eq.~(\ref{eqn:SpectralFunction}).]
Noting that $N(\omega)\sim k_BT/\hbar\omega$ for $\omega\sim0$,
one sees that the only spectral functions that are linear (Ohmic)
for small positive $\omega$,
\begin{equation}
\label{eqn:Ohmic}
{\mit\Gamma}_{\alpha\alpha'}(\omega)
\sim\eta_{\alpha\alpha'}\omega
=C_\alpha C_{\alpha'}^*\eta\omega,
\end{equation}
yield nontrivial values for
$\lim_{\lambda\to0}{\mit\Gamma}_{\alpha\alpha'}^\pm%
(\lambda^2\omega_{\alpha\alpha'})$.
For $\omega_{\alpha\alpha'}>0$, the zeroth order of the expansion
in $\lambda^2\hbar\omega_{\alpha\alpha'}/k_BT$,
\begin{equation}
\label{eqn:Expansion}
{\mit\Gamma}_{\alpha\alpha'}^\pm(\lambda^2\omega_{\alpha\alpha'})
/(k_BT/\hbar)
=\eta_{\alpha\alpha'}
+O(\lambda^2\hbar\omega_{\alpha\alpha'}/k_BT),
\end{equation}
survives in the stochastic limit $\lambda\to0$, so that
\begin{equation}
\lim_{\lambda\to0}
{\mit\Gamma}_{\alpha\alpha'}^\pm(\lambda^2\omega_{\alpha\alpha'})
=\left\{
\begin{array}{cl}
\medskip
\displaystyle
\frac{k_BT}{\hbar}\eta_{\alpha\alpha'}&
(\omega_{\alpha\alpha'}>0)\\
\medskip
\displaystyle
\frac{k_BT}{2\hbar}\eta_{\alpha\alpha'}&
(\omega_{\alpha\alpha'}=0)\\
\displaystyle
0&
(\omega_{\alpha\alpha'}<0).
\end{array}
\right.
\end{equation}
The case $\omega_{\alpha\alpha'}=0$ is treated separately since
${\mit\Gamma}_{\alpha\alpha'}^\pm(\omega)$ is discontinuous at
$\omega=0$, and ${\mit\Gamma}_{\alpha\alpha'}^\pm(0)$ is defined
here by ${\mit\Gamma}_{\alpha\alpha'}^\pm(0)%
=[{\mit\Gamma}_{\alpha\alpha'}^\pm(0^+)%
+{\mit\Gamma}_{\alpha\alpha'}^\pm(0^-)]/2$ for convention.

The correlation functions~(\ref{eqn:Correlation}) are similar to
their counterparts in the usual treatment for weak
damping~\cite{ref:SLA,ref:SLAinSpinBoson}, which are reproduced
by the replacement $\lambda^2\omega_\alpha\to\omega_\alpha$ in
Eqs.~(\ref{eqn:Correlation}).
There is, however, a significant difference between them:
In the usual stochastic limit for weak damping, the rapidly
oscillating factor
$e^{i(\omega_\alpha-\omega_{\alpha'})\tau/\lambda^2}$ extracts
the $\omega_\alpha=\omega_{\alpha'}$ contribution only and makes
the ``rotating-wave approximation''
exact~\cite{ref:ExactDiagonalization}, while in the present case,
the rotating-wave approximation is not applicable in general.

Now let us derive the Heisenberg equations for the dressed spin
system operators in the bath,
\begin{equation}
\label{eqn:DressedSpinOperator}
{\mathcal D}_\alpha(\tau)
=\tr_B\!\left(
\rho_B\,{\mathcal U}_I^\dagger(\tau)D_\alpha
e^{i\omega_\alpha\tau}{\mathcal U}_I(\tau)
\right),
\end{equation}
from the Schr\"odinger
equation~(\ref{eqn:SchroedingerEqInStochasticLimit}).
The above set of equations yields the Bloch equation and is
immediately translated into the master equation.
Differentiating both sides of Eq.~(\ref{eqn:DressedSpinOperator})
with respect to $\tau$, and applying the Schr\"odinger
equation~(\ref{eqn:SchroedingerEqInStochasticLimit}) to the
time derivative of ${\mathcal U}_I(\tau)$, we will obtain the
equations for ${\mathcal D}_\alpha(\tau)$ after computing the
partial trace with the thermal state $\rho_B$ in
Eq.~(\ref{eqn:DressedSpinOperator}).
Since we are working in the operator formalism, we rely upon the
techniques of Thermo Field Dynamics (TFD)~\cite{ref:TFDTextbook}
for the evaluation of the trace with the thermal state.
In the language of TFD, the thermal state is represented by
``thermal vacuum'' $|\theta\rangle$.
It is annihilated by the annihilation operator $\xi_k$ and its
tilde conjugate $\tilde{\xi}_k$, which are related to the
operator $a_k$ by
\begin{equation}
\label{eqn:ThermalBogoliubovTr}
a_k
=\sqrt{\,1+N(\omega_k)}\xi_k
+\sqrt{N(\omega_k)}\tilde{\xi}_k^\dagger.
\end{equation}
Thus the key formulas for the derivation of the Heisenberg
equations for ${\mathcal D}_\alpha(\tau)$ are the commutation
relations between ${\mathcal U}_I(\tau)$ and the operators
$\chi_\alpha(\tau)$, $\tilde{\chi}_\alpha(\tau)$, defined as the
stochastic limit of the operators
${\mit\Xi}_\alpha(\tau/\lambda^2)/\lambda$ and
$\tilde{\mit\Xi}_\alpha(\tau/\lambda^2)/\lambda$:
\begin{mathletters}
\begin{equation}
{\mit\Xi}_\alpha(t)
=C_\alpha\int dk\sqrt{\,1+N(\omega_k)}g_k\xi_k
e^{-i(\omega_k-\lambda^2\omega_\alpha)t},
\end{equation}
\begin{equation}
\tilde{\mit\Xi}_\alpha(t)
=C_\alpha^*\int dk\sqrt{N(\omega_k)}g_k^*\tilde{\xi}_k
e^{i(\omega_k-\lambda^2\omega_\alpha)t}.
\end{equation}
\end{mathletters}
Notice that they are related to $b_\alpha(\tau)$ by
$b_\alpha(\tau)=\chi_\alpha(\tau)%
+\tilde{\chi}_\alpha^\dagger(\tau)$.

The commutation relations enable us to make normal ordering and
to evaluate the partial trace.
They are calculated in a way similar to that in the usual
stochastic limit for weak
damping~\cite{ref:SLA,ref:SLAinSpinBoson}:
Observe the commutation relations between
${\mit\Xi}_\alpha(\tau/\lambda^2)/\lambda$,
$\tilde{\mit\Xi}_\alpha(\tau/\lambda^2)/\lambda$ and the
iterative solution
\begin{equation}
U_I(\tau/\lambda^2)
=1-\frac{i}{\hbar}\int_0^\tau d\tau'
\frac{1}{\lambda}V_I(\tau'/\lambda^2)U_I(\tau'/\lambda^2),
\end{equation}
and then take the stochastic limit.
The key commutation relations are obtained as
\begin{mathletters}
\label{eqn:Commutator}
\begin{eqnarray}
&&[\chi_\alpha(\tau),{\mathcal U}_I(\tau)]
=-\sum_{\alpha'}\lim_{\lambda\to0}\Bigl(
i{\mit\Sigma}_{\alpha\alpha'}^+(\lambda^2\omega_{\alpha\alpha'})
\Bigr)\nonumber\\
&&\qquad\qquad\qquad\qquad\qquad\ %
{}\times e^{i(\omega_\alpha-\omega_{\alpha'})\tau}
D_{\alpha'}^\dagger{\mathcal U}_I(\tau),
\end{eqnarray}
\begin{eqnarray}
&&[\tilde{\chi}_\alpha(\tau),{\mathcal U}_I(\tau)]
=\sum_{\alpha'}\lim_{\lambda\to0}\Bigl(
i{\mit\Sigma}_{\alpha\alpha'}^-(\lambda^2\omega_{\alpha\alpha'})
\Bigr)^*\nonumber\\
&&\qquad\qquad\qquad\qquad\qquad\ %
{}\times e^{-i(\omega_\alpha-\omega_{\alpha'})\tau}
D_{\alpha'}{\mathcal U}_I(\tau).
\end{eqnarray}
\end{mathletters}
Here ${\mit\Sigma}_{\alpha\alpha'}^\pm(\omega)$ is the
self-energy function, given by
\begin{mathletters}
\begin{eqnarray}
{\mit\Sigma}_{\alpha\alpha'}^\pm(\omega)
&=&-i\int_0^\infty dt\int_0^\infty\frac{d\omega'}{2\pi}
{\mit\Gamma}_{\alpha\alpha'}^\pm(\omega')
e^{-i(\omega'-\omega)t}\\
&=&\int_0^\infty\frac{d\omega'}{2\pi}
{\mit\Gamma}_{\alpha\alpha'}^\pm(\omega')
\frac{1}{\omega-\omega'+i0^+}\\
&=&{\mit\Delta}_{\alpha\alpha'}^\pm(\omega)
-\frac{i}{2}{\mit\Gamma}_{\alpha\alpha'}^\pm(\omega),
\end{eqnarray}
whose imaginary part ${\mit\Gamma}_{\alpha\alpha'}^\pm(\omega)$
will give the decay rate, and real part
\begin{equation}
{\mit\Delta}_{\alpha\alpha'}^\pm(\omega)
={\mathcal P}\int_0^\infty\frac{d\omega'}{2\pi}
{\mit\Gamma}_{\alpha\alpha'}^\pm(\omega')
\frac{1}{\omega-\omega'}
\end{equation}
\end{mathletters}
will contribute to the energy shift in the following Bloch
equation or in the master equation.
The counterparts in the usual stochastic limit for weak
damping~\cite{ref:SLA,ref:SLAinSpinBoson} are again reproduced
from Eqs.~(\ref{eqn:Commutator}) by the same replacement
$\lambda^2\omega_\alpha\to\omega_\alpha$ as that for the
correlation functions~(\ref{eqn:Correlation}).

By making use of the commutation
relations~(\ref{eqn:Commutator}), the Bloch equation for rapidly
decaying spin system is obtained:
\end{multicols}\widetext
\begin{equation}
\label{eqn:BlochEq}
\frac{d}{d\tau}\left(
\begin{array}{cc}
\medskip
{\mathcal D}_+(\tau)\\
\medskip
{\mathcal D}_0(\tau)\\
{\mathcal D}_-(\tau)
\end{array}
\right)=\left(
\begin{array}{ccc}
\medskip
-(\tilde{\mit\Delta}^2+2\tilde{\varepsilon}^2)\gamma^\theta\!/2
+i{\mit\Omega}_0&
\tilde{\varepsilon}\tilde{\mit\Delta}\gamma^\theta\!/2&
\tilde{\mit\Delta}^2\gamma^\theta\!/2\\
\medskip
\tilde{\varepsilon}\tilde{\mit\Delta}\gamma^\theta&
-\tilde{\mit\Delta}^2\gamma^\theta&
\tilde{\varepsilon}\tilde{\mit\Delta}\gamma^\theta\\
\tilde{\mit\Delta}^2\gamma^\theta\!/2&
\tilde{\varepsilon}\tilde{\mit\Delta}\gamma^\theta\!/2&
-(\tilde{\mit\Delta}^2+2\tilde{\varepsilon}^2)\gamma^\theta\!/2
-i{\mit\Omega}_0
\end{array}
\right)\left(
\begin{array}{cc}
\medskip
{\mathcal D}_+(\tau)\\
\medskip
{\mathcal D}_0(\tau)\\
{\mathcal D}_-(\tau)
\end{array}
\right),
\end{equation}
\wideeqend
\begin{multicols}{2}\narrowtext\noindent
where $\tilde{\varepsilon}=\varepsilon/\hbar{\mit\Omega}_0$,
$\tilde{\mit\Delta}={\mit\Delta}/\hbar{\mit\Omega}_0$, and
\begin{equation}
\label{eqn:DampingFactor}
\gamma^\theta=2\eta k_BT/\hbar.
\end{equation}
Through the relations
$\langle\pm|\rho_S(\tau)|\mp\rangle=\tr_S[\rho_S{\mathcal
D}_\mp(\tau)]$ and
$\langle\pm|\rho_S(\tau)|\pm\rangle=\{1\pm\tr_S[\rho_S{\mathcal
D}_0(\tau)]\}/2$, this gives the master equation for the density
operator of the spin system, $\rho_S(\tau)=\tr_B\rho_{\mathrm
tot}(\tau)$,
\begin{equation}
\label{eqn:MasterEq}
\frac{d}{d\tau}\rho_S(\tau)
=-\frac{i}{\hbar}[H_S,\rho_S(\tau)]
-\frac{\gamma^\theta}{4}
\bm{[}\sigma_z,[\sigma_z,\rho_S(\tau)]\bm{]}.
\end{equation}
Note that, for the model considered here, the energy shift
disappears in the stochastic limit.

The key formulas for weak damping and those for rapid decay are
quite similar: The formulas for weak damping are reproduced by
the formal replacement $\lambda^2\omega_\alpha\to\omega_\alpha$
in those for rapid decay in Eqs.~(\ref{eqn:Correlation})
and~(\ref{eqn:Commutator}).
However, in the case of weak damping where the total Hamiltonian
is given by Eqs.~(\ref{eqn:WeakDampingHamiltonian})
and~(\ref{eqn:SpinBosonModel}), the Bloch equation is much
simpler than that for rapid decay~(\ref{eqn:BlochEq}).
The matrix in Eq.~(\ref{eqn:BlochEq}) for weak damping is
diagonal (or decoupled), and furthermore, the damping
coefficients of ${\mathcal D}_\pm(\tau)$ and ${\mathcal
D}_0(\tau)$ are given by
$\gamma_D^\theta=\tilde{\mit\Delta}^2\gamma^\theta\!/2$ and
$\gamma_R^\theta=\tilde{\mit\Delta}^2\gamma^\theta$,
respectively, but with
$\gamma^\theta={\mit\Gamma}({\mit\Omega}_0)\coth%
(\hbar{\mit\Omega}_0/2k_BT)$ instead of that for rapid
decay~(\ref{eqn:DampingFactor})~\cite{ref:SLAinSpinBoson}, i.e.,
the relation $\tau_D=2\tau_R$ always holds between the
decoherence time $\tau_D=(\gamma_D^\theta)^{-1}$ and the thermal
relaxation time $\tau_R=(\gamma_R^\theta)^{-1}$, while it does
not always hold in spin-relaxation experiments.
The phenomenological Bloch equation~\cite{ref:BlochEq} is given
two independent time scales $\tau_R$ and $\tau_D$.
The simplicity of the weak damping formulas is due to the
rotating-wave approximation, arising from the frequent
oscillations of the spin system
$e^{i(\omega_\alpha-\omega_{\alpha'})\tau/\lambda^2}$, which
suppresses many terms in the
equations~\cite{ref:ExactDiagonalization}.
The Bloch equation~(\ref{eqn:BlochEq}) for rapidly decaying
systems, on the other hand, contains a richer variety.
Between the two decay constants, which are the real parts of the
eigenvalues of the matrix in Eq.~(\ref{eqn:BlochEq}), there is no
trivial relation as that for weak damping.
Even more, there is a case, depending on the parameters, where
three decay constants (three real eigenvalues) exist.
Details of these features will be reported in elsewhere.

It is easy to show that the master equation~(\ref{eqn:MasterEq})
has the unique thermal equilibrium state $\rho_{\mathrm eq}$,
which is proportional to $1$: $\rho_S(\tau)\to\rho_{\mathrm
eq}\propto1$ as $\tau\to\infty$.
$\rho_{\mathrm eq}$ is nothing but the thermal state at
infinitely high temperature.
This is because the master equation~(\ref{eqn:MasterEq}) is valid
for the situation $\lambda^2\hbar{\mit\Omega}_0/k_BT\ll1$.
Remember the expansion~(\ref{eqn:Expansion}).
The stochastic limit approximation is a method to extract the
minimal dissipative dynamics from the full of it.
It picks only the processes where one boson is emitted or
absorbed, and neglects the higher-order contributions~\cite{%
ref:PascazioFacchiNonexponential-VanHoveLimit}.
As for the stochastic limit approximation for rapid decay, one
can say, in addition, that it extracts the dissipative dynamics
\textit{at a much higher temperature} than the characteristic
energy scales of the system.

The master equation~(\ref{eqn:MasterEq}) derived through the
stochastic limit approximation for rapid decay is worthy of note.
It is of the Lindblad form, which ensures the conservation and
positivity of probability~\cite{ref:LindbladCommunMathPhys1976}.
For the case
of weak damping, many authors derived master equations of the
Lindblad form, while for the case of rapid decay, much care is
required~\cite{ref:QuantumNoise,ref:NonRotatingWaveMasterEq}.
The master equation derived by Caldeira and
Leggett~\cite{ref:CaldeiraLeggettPhysica}, for example, is not of
the Lindblad form~\cite{ref:QuantumNoise,ref:NonLindblad,%
ref:NonRotatingWaveMasterEq}.
Such master equations lead, in certain cases, to unphysical
results~\cite{ref:NonRotatingWaveMasterEq}.

It is interesting to look at the master equation for the model
whose total Hamiltonian is given by
Eq.~(\ref{eqn:RapidDecayHamiltonian}) with
\begin{mathletters}
\label{eqn:CaldeiraLeggettModel}
\begin{equation}
H_S=\frac{1}{2M}p^2+\frac{1}{2}M\!{\mit\Omega}_0^2x^2,
\end{equation}
\begin{equation}
V=-i\hbar\sqrt{\frac{M\!{\mit\Omega}_0}{\hbar}}x
\int dk\left(g_ka_k-g_k^*a_k^\dagger\right),
\end{equation}
\end{mathletters}
and the same $H_B$ as in Eqs.~(\ref{eqn:BathHamiltonian}).
The master equation derived through the stochastic limit
approximation for rapid decay illustrated here reads
\begin{mathletters}
\label{eqn:MasterEqForCaldeiraLeggettModel}
\begin{equation}
\label{eqn:LindbladMasterEqForCaldeiraLeggettModel}
\frac{d}{d\tau}\rho_S(\tau)
=-\frac{i}{\hbar}[H_S^\theta,\rho_S(\tau)]
-\frac{M\!{\mit\Omega}_0\eta k_BT}{2\hbar^2}
\bm{[}x,[x,\rho_S(\tau)]\bm{]},
\end{equation}
where $H_S^\theta$ is the renormalized Hamiltonian with the
renormalized frequency
\begin{equation}
\label{eqn:RenormalizedFrequency}
{\mit\Omega}_R^2
={\mit\Omega}_0^2
-2{\mit\Omega}_0\int_0^\infty \frac{d\omega}{2\pi}
\frac{\mit\Gamma(\omega)}{\omega}.
\end{equation}
\end{mathletters}
The master
equation~(\ref{eqn:LindbladMasterEqForCaldeiraLeggettModel}) is
of the Lindblad form since it lacks the term
$-(i{\mit\Omega}_0\eta/4\hbar)[x,\{p,\rho_S(\tau)\}]$ which is
contained in the one derived by Caldeira and
Leggett~\cite{ref:CaldeiraLeggettPhysica}.
This term may be neglected, when compared to the last term of the
master
equation~(\ref{eqn:LindbladMasterEqForCaldeiraLeggettModel}) if
the temperature $T$ is high enough.
The same situation is found in the spin-boson model.
The master equation derived by Munro and Gardiner in
Ref.~\cite{ref:NonRotatingWaveMasterEq}, which is again not of
the Lindblad form, is reduced to the Lindblad form master
equation~(\ref{eqn:MasterEq}) in the high-temperature limit.

The model~(\ref{eqn:CaldeiraLeggettModel}) treated above is
exactly solvable~\cite{ref:Ullersma,ref:GrabertSchrammIngold,%
ref:HuPazQBM,ref:ExactDiagonalization}, and it is possible to
write down an exact master equation for the reduced density
operator~\cite{ref:HuPazQBM}.
Let us finally confirm the validity of the master
equation~(\ref{eqn:MasterEqForCaldeiraLeggettModel}) derived
through the method presented here by comparing it with the exact
one.
The master equation, which is exact for an arbitrary $\lambda$,
is obtained through the method sketched in the Appendix and reads
\begin{mathletters}
\label{eqn:ExactMasterEq}
\begin{eqnarray}
&&i\hbar\frac{d}{d\tau}\rho_S(\tau)
=[H_S^\theta(\tau),\rho_S(\tau)]
-iD_{xx}(\tau)\bm{[}x,[x,\rho_S(\tau)]\bm{]}
\nonumber\\
&&\qquad{}-2iD_{xp}(\tau)\bm{[}x,[p,\rho_S(\tau)]\bm{]}
+{\mit\Gamma}_{xp}(\tau)[x,\{p,\rho_S(\tau)\}],\nonumber\\
\end{eqnarray}
where
\begin{equation}
H_S^\theta(\tau)
=\frac{1}{2M}p^2
+\frac{1}{2}M\!{\mit\Omega}_R^2(\tau)x^2.
\end{equation}
\end{mathletters}
The time-dependent coefficients ${\mit\Omega}_R(\tau)$,
$D_{xx}(\tau)$, $D_{xp}(\tau)$, and ${\mit\Gamma}_{xp}(\tau)$ are
given by Eqs.~(\ref{eqn:ExactCoefficients}).
They are reduced in the stochastic limit $\lambda\to0$, as shown
in the Appendix, to
\begin{equation}
\label{eqn:CoefficientsInTheStochasticLimit}
\begin{array}{c}
\bigskip
\displaystyle
{\mit\Omega}_R^2(\tau)\to{\mit\Omega}_R^2,\quad
D_{xx}(\tau)\to\frac{M\!{\mit\Omega}_0\eta k_BT}{2\hbar},\\
\displaystyle
D_{xp}(\tau)\to0,\quad
{\mit\Gamma}_{xp}(\tau)\to0,
\end{array}
\end{equation}
and the master
equation~(\ref{eqn:MasterEqForCaldeiraLeggettModel}) is thus
reproduced from the exact one~(\ref{eqn:ExactMasterEq}), which
shows the consistency of our method.

We have discussed the stochastic limit approximation for rapid
decay and extended its applicability.
Although some mathematical issues, e.g., the question of the
convergence of the operators, remain unaddressed, the framework
presented here is useful for practical calculations.

The authors acknowledge useful and helpful discussions with
Professor~I.~Ohba.
They also thank H.~Nakazato and S.~Pascazio for critical reading
of the manuscript and enlightening comments, P.~Facchi for
fruitful discussions, and M.~Miyamoto for advice.
This work was supported by a Grant-in-Aid for JSPS Research
Fellows.

\appendix
\section*{}
Here we briefly illustrate a derivation of the exact master
equation~(\ref{eqn:ExactMasterEq}) for the
model~(\ref{eqn:RapidDecayHamiltonian}) with $H_S$, $V$ in
Eqs.~(\ref{eqn:CaldeiraLeggettModel}), and $H_B$ in
Eqs.~(\ref{eqn:BathHamiltonian}), and show that the stochastic
limit of it reproduces the master
equation~(\ref{eqn:MasterEqForCaldeiraLeggettModel}).

The derivation of the exact master
equation~(\ref{eqn:ExactMasterEq}) is found in
Ref.~\cite{ref:HuPazQBM}, which is based on the influence
functional approach proposed by Feynman and
Vernon~\cite{ref:FeynmanVernon}.
The path integration of the time-evolution kernel for the reduced
density matrix in the position representation,
$J(x_f,x_f',\tau_f|x_i,x_i',\tau_i)$ [with which the reduced
density matrix is given by $\langle
x|\rho_S(\tau)|x'\rangle=\int_{-\infty}^\infty
dx_0\int_{-\infty}^\infty dx_0'\,J(x,x',\tau|x_0,x_0',0)\langle
x_0|\rho_S|x_0'\rangle$], is carried out without any
approximation to give~\cite{ref:GrabertSchrammIngold,%
ref:HuPazQBM}
\begin{eqnarray}
&&J(x_f,x_f',\tau|x_i,x_i',0)
=\frac{1}{2\pi\hbar}|{\mit\Lambda}_{if}(\tau)|\nonumber\\
&&\quad{}\times\exp\biggl[
\sum_{k,l}\left(
\frac{i}{\hbar}
{\mit\Delta}_k{\mit\Lambda}_{kl}(\tau){\mit\Sigma}_l
-\frac{1}{\hbar}
{\mit\Delta}_k{\mit\Theta}_{kl}(\tau){\mit\Delta}_l
\right)
\biggr],
\label{eqn:ExactSolution}
\end{eqnarray}
where ${\mit\Sigma}_k=(x_k+x_k')/2$, ${\mit\Delta}_k=x_k-x_k'$
($k=\{f,i\}$),
\begin{mathletters}
\begin{equation}
{\mit\Lambda}_{ff}(\tau)={\mit\Lambda}_{ii}(\tau)
=M\frac{d}{d\tau}\ln|F(\tau)|,
\end{equation}
\begin{equation}
{\mit\Lambda}_{fi}(\tau)
=\lambda^2MF(\tau)\frac{d^2}{d\tau^2}\ln|F(\tau)|,
\quad
{\mit\Lambda}_{if}(\tau)=-\frac{M}{\lambda^2F(\tau)},
\end{equation}
\end{mathletters}
and
\begin{mathletters}
\begin{equation}
{\mit\Theta}_{kl}(\tau)
=\frac{1}{2}\int_0^\tau d\tau_1\int_0^\tau d\tau_2\,
c_k(\tau,\tau_1)\nu(\tau_1-\tau_2)c_l(\tau,\tau_2),
\end{equation}
\begin{equation}
c_f(\tau,\tau')
=\lambda^2\dot{F}(\tau')
-\left(\frac{d}{d\tau}\ln|F(\tau)|\right)\lambda^2F(\tau'),
\end{equation}
\begin{equation}
c_i(\tau,\tau')=\frac{F(\tau')}{F(\tau)},
\end{equation}
\begin{equation}
\nu(\tau)
=M\!{\mit\Omega}_0\int_0^\infty\frac{d\tilde{\omega}}{2\pi}
{\mit\Gamma}(\lambda^2\tilde{\omega})
\coth\frac{\lambda^2\hbar\tilde{\omega}}{2k_BT}
\cos\tilde{\omega}\tau.
\end{equation}
\end{mathletters}
$F(\tau)$ is the solution of the integrodifferential equation
\begin{mathletters}
\begin{equation}
\ddot{F}(\tau)+{\mit\Omega}_0^2F(\tau)
+\frac{2}{M}\int_0^\tau d\tau'\mu(\tau-\tau')F(\tau')=0,
\end{equation}
\begin{equation}
\mu(\tau)
=-M\!{\mit\Omega}_0\int_0^\infty\frac{d\tilde{\omega}}{2\pi}
{\mit\Gamma}(\lambda^2\tilde{\omega})\sin\tilde{\omega}\tau
\end{equation}
\end{mathletters}
with the initial condition $F(0)=0$, $\dot{F}(0)=1$, and is given
by an inverse Laplace transformation
\begin{mathletters}
\begin{equation}
\lambda^2F(\tau)
=\int_C\frac{d\tilde{s}}{2\pi i}
\frac{1}{%
\tilde{s}^2+{\mit\Omega}_0^2
+2\hat{\mu}(\lambda^2\tilde{s})/M}e^{\tilde{s}\tau
},
\end{equation}
\begin{equation}
\hat{\mu}(s)
=-M\!{\mit\Omega}_0\int_0^\infty\frac{d\omega}{2\pi}
{\mit\Gamma}(\omega)\frac{\omega}{s^2+\omega^2}.
\end{equation}
\end{mathletters}
$C$ is the so-called Bromwich path.
Note that time $t$ has already been rescaled to $\tau=\lambda^2t$
(and accordingly, the integration variable $\omega$ to
$\tilde{\omega}=\omega/\lambda^2$), but $\lambda$ is so far
finite, and the solution~(\ref{eqn:ExactSolution}) is exact for
an arbitrary $\lambda$.

The exact master equation~(\ref{eqn:ExactMasterEq}) is then
derived by differentiating the kernel~(\ref{eqn:ExactSolution})
with respect to time $\tau$, replacing ${\mit\Delta}_i$ and
${\mit\Sigma}_i$ with linear combinations of
$\partial/\partial{\mit\Delta}_f$,
$\partial/\partial{\mit\Sigma}_f$, ${\mit\Delta}_f$, and
${\mit\Sigma}_f$~\cite{ref:HuPazQBM}, multiplying by the initial
density $\langle x_i|\rho_S|x_i'\rangle$, and integrating over
the initial coordinates $x_i$, $x_i'$.
The coefficients of the master equation~(\ref{eqn:ExactMasterEq})
are given by
\begin{mathletters}
\label{eqn:ExactCoefficients}
\begin{equation}
{\mit\Omega}_R^2(\tau)
=\frac{1}{M}{\mit\Lambda}_{ff}
\frac{d}{d\tau}\ln\left|
\frac{{\mit\Lambda}_{fi}}{{\mit\Lambda}_{ff}}
\right|,
\end{equation}
\begin{eqnarray}
D_{xx}(\tau)
&=&4{\mit\Gamma}_{xp}(\tau)
\left(
{\mit\Theta}_{ff}
-\frac{{\mit\Theta}_{fi}}{{\mit\Lambda}_{if}}{\mit\Lambda}_{ff}
\right)\nonumber\\
&&{}+\dot{\mit\Theta}_{ff}
-\frac{2\dot{\mit\Theta}_{fi}}{{\mit\Lambda}_{if}}
{\mit\Lambda}_{ff}
+\frac{\dot{\mit\Theta}_{ii}}{{\mit\Lambda}_{if}^2}
{\mit\Lambda}_{ff}^2,
\end{eqnarray}
\begin{eqnarray}
D_{xp}(\tau)
&=&\frac{2{\mit\Theta}_{fi}}{{\mit\Lambda}_{if}}
{\mit\Gamma}_{xp}(\tau)
+\frac{1}{M}\left(
{\mit\Theta}_{ff}-\frac{{\mit\Theta}_{fi}}{{\mit\Lambda}_{if}}
{\mit\Lambda}_{ff}
\right)\nonumber\\
&&{}+\frac{\dot{\mit\Theta}_{fi}}{{\mit\Lambda}_{if}}
-\frac{\dot{\mit\Theta}_{ii}}{{\mit\Lambda}_{if}^2}
{\mit\Lambda}_{ff},
\end{eqnarray}
and
\begin{equation}
{\mit\Gamma}_{xp}(\tau)=-\frac{1}{2}\left(
\frac{d}{d\tau}\ln|{\mit\Lambda}_{fi}|
+\frac{1}{M}{\mit\Lambda}_{ff}
\right).
\end{equation}
\end{mathletters}

Now we observe that the master
equation~(\ref{eqn:MasterEqForCaldeiraLeggettModel}) is
reproduced from the exact one~(\ref{eqn:ExactMasterEq}) in the
stochastic limit $\lambda\to0$.
First notice that the function $\lambda^2F(\tau)$ is evaluated in
the limit as
\begin{equation}
\lambda^2F(\tau)
\to\int_C\frac{d\tilde{s}}{2\pi i}
\frac{1}{\tilde{s}^2+{\mit\Omega}_R^2}e^{\tilde{s}\tau}
=\frac{1}{{\mit\Omega}_R}\sin{\mit\Omega}_R\tau
\end{equation}
with the renormalized frequency ${\mit\Omega}_R$ given in
Eq.~(\ref{eqn:RenormalizedFrequency}).
${\mit\Lambda}_{kl}(\tau)$ are hence given by
\begin{mathletters}
\label{eqn:LambdaInStochasticLimit}
\begin{equation}
{\mit\Lambda}_{ff}(\tau)
={\mit\Lambda}_{ii}(\tau)
\to M\!{\mit\Omega}_R\cot{\mit\Omega}_R\tau,
\end{equation}
\begin{equation}
{\mit\Lambda}_{fi}(\tau),\ {\mit\Lambda}_{if}(\tau)
\to-\frac{M\!{\mit\Omega}_R}{\sin{\mit\Omega}_R\tau},
\end{equation}
\end{mathletters}
and ${\mit\Theta}_{kl}(\tau)$ by
\begin{mathletters}
\label{eqn:ThetaInStochasticLimit}
\begin{equation}
{\mit\Theta}_{ff}(\tau),\ {\mit\Theta}_{ii}(\tau)
\to\frac{M\!{\mit\Omega}_0\eta k_BT}{\hbar}
\frac{%
{\mit\Omega}_R\tau-\sin{\mit\Omega}_R\tau\cos{\mit\Omega}_R\tau%
}{4{\mit\Omega}_R\sin^2\!{\mit\Omega}_R\tau},
\end{equation}
\begin{eqnarray}
&&{\mit\Theta}_{fi}(\tau)={\mit\Theta}_{if}(\tau)\nonumber\\
&&\qquad\to-\frac{M\!{\mit\Omega}_0\eta k_BT}{\hbar}
\frac{%
{\mit\Omega}_R\tau\cos{\mit\Omega}_R\tau-\sin{\mit\Omega}_R\tau%
}{4{\mit\Omega}_R\sin^2\!{\mit\Omega}_R\tau}.
\end{eqnarray}
\end{mathletters}
It is easy to see that the
substitution~(\ref{eqn:LambdaInStochasticLimit}) and
(\ref{eqn:ThetaInStochasticLimit}) brings us with those constants
given in Eqs.~(\ref{eqn:CoefficientsInTheStochasticLimit}), and
the exact master equation~(\ref{eqn:ExactMasterEq}) is reduced to
the master equation~(\ref{eqn:MasterEqForCaldeiraLeggettModel}).


\end{multicols}
\end{document}